\documentclass[11pt]{article}
\usepackage{latexsym}
\usepackage[dvips]{epsfig}
\usepackage{color}
\usepackage{amsmath}
\usepackage{amssymb}
\usepackage{amsthm}
\usepackage{amsfonts}
\usepackage{mathrsfs}
\usepackage{graphics}
\usepackage{graphicx}
\usepackage{curves}
\usepackage{subfigure}
\usepackage{caption}
\usepackage{framed}
\usepackage{fancybox}
\usepackage{epic,eepic}
\usepackage{algorithm}
\usepackage{algorithmic}
\usepackage{longtable}
\usepackage{multirow}
\usepackage{pdflscape}
\usepackage{hhline}
\usepackage{tikz}
\usepackage{adjustbox}
\usepackage{geometry}
\usepackage{authblk}
\usepackage{fancyhdr}
\usepackage{lipsum}

\newtheorem{thm}{Theorem}
\newtheorem{lem}{Lemma}
\newtheorem{prop}{Property}
\newtheorem*{defn*}{Definition}

\def\proof{\pn {Proof.} }

\def\endproof{\hfill $\Box$ \vskip .5cm}

\def\pn{\par\smallskip\noindent}

\begin{document}

\title{A DFS Algorithm for Maximum Matchings in General Graphs}

\author[label1]{Tony T. Lee
\footnote{Tony T. Lee is with the Guangdong Provincial Key Laboratory of Future Networks of Intelligence, The Chinese University of Hong Kong (Shenzhen).}, \hspace{6mm}
Bojun Lu$^{\ddagger}$, \hspace{6mm}
Hanli Chu$^{\mathsection}$\\
\vspace{6mm}
\emph{The Chinese University of Hong Kong (Shenzhen), Shenzhen 518000 China
\footnote{This work was supported in part by the National Science Foundation of China under Grant 61571288, in part by the Shenzhen Science and Technology Innovation Committee under Grant JCYJ20180508162604311, in part by the National Key Research and Development Program of China under Grant 2018YFB1800803, and in part by the Guangdong Provincial Key Laboratory of Future Networks of Intelligence.}\\
\vspace{3mm}
$^{*}$tonylee@cuhk.edu.cn, \hspace{2mm}
$^{\ddagger}$bojunlu@cuhk.edu.cn, \hspace{2mm}
$^{\mathsection}$117010044@link.cuhk.edu.cn
}}

\date{\today}

\maketitle

\begin{abstract}
\noindent
In this paper, we propose a depth-first search (DFS) algorithm for searching maximum matchings in general graphs. Unlike blossom shrinking algorithms, which store all possible alternative alternating paths in the super-vertices shrunk from blossoms, the newly proposed algorithm does not involve blossom shrinking. The basic idea is to deflect the alternating path when facing blossoms. The algorithm maintains detour information in an auxiliary stack to minimize the redundant data structures. A benefit of our technique is to avoid spending the time on shrinking and expanding blossoms.
This DFS algorithm can determine a maximum matching of a general graph with $m$ edges and $n$ vertices in $O\left(mn\right)$ time with space complexity $O\left(n\right)$.

\vspace{1cm}

\end{abstract}

\section{Introduction}\label{sec:01}

\hspace{4mm}
The maximum matching in an undirected graph is a set of disjoint edges that has the maximum cardinality. Finding a maximum matching is a fundamental problem in combinatorial optimization \cite{lovasz2009matching}. It has wide applications in the development of graph theory and computer science \cite{cormen2009introduction}. In a bipartite graph with $n$ vertices and $m$ edges, finding a maximum matching problem is solved by the Hopcroft-Karp algorithm in time $O((m+n)\sqrt{n})$ time \cite{hopcroft1973n}. For constructing a maximum matching in a general graph, the blossom shrinking algorithm proposed by Edmonds in \cite{Edmonds:1965} is the first polynomial algorithm that runs in time $O\left(n^4\right)$. The complexity of this algorithm has been improved from $O\left(n^4\right)$ to $O\left(n^3\right)$ by Gabow \cite{Gabow:1976}, and further reduced to $O\left(mn\right)$ by Gabow and Tarjan \cite{GabowTarjan:1985} for a graph with $n$ vertices and $m$ edges. The best-known algorithm is given by Micali and Vazirani \cite{MicaliVazirani:1980} and \cite{vazirani1994theory} that runs in $O\left(m\sqrt{n}\right)$, but it is rather difficult to understand and too complex for efficient implementation. Almost all these algorithms follow Edmonds' idea of shrinking blossoms \cite{tarjan1983data}, which requires data structures to represent blossoms, and the time spent on shrinking and expanding blossoms.

\hspace{4mm}
The main contribution of this paper is to present a DFS maximum matching algorithm that does not involve blossom shrinking. The basic idea is to deflect the alternating path when a blossom, or odd cycle, is formed. This deflection algorithm adopts two stacks, one is a directional alternating path, and the other one is an ordered list of edges to maintain detour information. The two stacks interact with each other to grow or to prune in the exploring process, until an augmenting path is identified or it confirms that no augmenting paths exist.
Unlike the Edmonds' algorithm, which is a breadth-first search (BFS) algorithm that stores all possible alternative alternating paths in the super-vertices shrunk from blossoms.
The deflection algorithm maintains such detour information in the sprout stack to minimize the redundant data structures. This newly proposed maximum matching algorithm can achieve a time complexity of $O(mn)$
with space complexity $O(n)$, because it avoids spending the time on shrinking and expanding blossoms.

\hspace{4mm}
The organization of this paper is as follows. In Section 2, we present the definitions of terminologies used in this paper. In Section 3, we describe the issue of parity conflicts arising from blossoms and illustrate our deflection method with some examples. This section is mainly expository in nature, and it compares the method of deflection versus shirking when blossoms occur.
In Section 4, we present a DFS algorithm to enumerate augmenting paths, and discuss the performance of this algorithm. Finally, we conclude this paper in Section 5.

\section{Preliminaries}\label{sec:02}

\begin{defn*}\emph{
An undirected graph $G(V,E)$ consists of a \textbf{vertex set} $V$ and an \textbf{edge set} $E$. An edge is an unordered pair of vertices $\{v,u\}$  and written as $e =  \langle v,u \rangle$. The number of vertices $n=|V(G)|$ is the \textbf{order} of $G$, and the number of edges $m=|E(G)|$ is the \textbf{size} of $G$.
}
\end{defn*}

\hspace{4mm}
Without loss of generality, we assume that the general graph $G$ under consideration is a simple graph without loops, multiple edges, or isolated vertices.

\begin{defn*}\emph{
A set $M \subseteq E$ is a \textbf{matching} if no two edges in $M$ have a vertex in common, or no vertex $v\in V $ is incident with more than one edge in $M$.
A matching of maximum cardinality is called a \textbf{maximum matching}.
A \textbf{perfect matching} of a graph $G$ is a matching which covers all vertices of $V$.
Relative to a matching $M$ in $G$, a vertex $v$ is called a \textbf{matched vertex}, or \textbf{covered vertex}, if it is incident to an edge in $M$. Otherwise, the vertex $v$ is called a \textbf{free vertex}. The set of $M$-matched vertices is denoted by $\partial(M)$, and the set of $M$-free vertices is denoted by
$\bar{\partial}(M)$.
Similarly, edges in $M$ are \textbf{matched edges}, while edges not in $M$ are \textbf{free edges}. Every matched vertex $v$ has a \textbf{mate}, the other endpoint of the matched edge.
}
\end{defn*}

\begin{defn*}\emph{
A \textbf{path} $P=\{\langle {v_1,v_2} \rangle, \langle {v_2,v_3} \rangle, \cdots, \langle {v_{l-1},v_{l}} \rangle\}$ is a sequence of edges, which alternately join a sequence of distinct vertices. The path $P$ can also sometimes be written as $P= v_1, v_2, \cdots, v_{l-1},v_{l}$.
A \textbf{cycle} is a path with an edge joining the first and last vertices.
Relative to a matching $M$ in $G$, an \textbf{$\boldsymbol{M}$-alternating path} $P$ is a path in which edges alternate between those in $M$ and those not in $M$, and it is called an \textbf{$\boldsymbol{M}$-augmenting path} if its endpoints $v_1$ and $v_{l}$ are both free, in which case $l$ must be even.
}
\end{defn*}

\hspace{4mm}
The following result shows that an $M$-augmenting path $P$ can enlarge the cardinality of $M$ by one.

\begin{lem}\label{lem:01}
{\emph{
If $P$ is an $M$-augmenting path relative to a matching $M$, then the symmetric difference defined by
\begin{align*}
M\oplus P := (M - P) \cup (P - M) = (M\cup P) - (M \cap P)
\end{align*}
is also a matching, and $|M\oplus P|=|M|+1$.
}}
\end{lem}

\hspace{4mm}
An immediate consequence of this result is the following theorem due to Berge \cite{Berge:1957} that characterizes maximum matchings.

\begin{thm}[Augmenting Path Theorem]\label{thm:01}
\emph{
A matching $M$ is maximum if and only if there is no $M$-augmenting path.
}
\end{thm}

\hspace{4mm}
Theorem~\ref{thm:01} implies that if a matching $M$ in a graph $G$ is not maximum, then there exists an $M$-augmenting path. This is the basis of all algorithms for determining maximum matchings in general graphs. The basic idea is to enlarge an existing matching $M$ by any $M$-augmenting path. Repeat the searching process until no augmenting paths exist anymore.

\hspace{4mm}
Suppose that $M$ is a matching in a graph $G(V,E)$, if we assign the red color to the edges in $M$ and the blue color to those edges not in $M$, then there is a one-to-one correspondence between the complex-colored graph and the matching $M$. Since all algorithms for finding maximum matching start with some existing matching, we will adopt the complex coloring method proposed in \cite{Lee:2013} and \cite{wang2018parallel} to initialize our maximum matching algorithm.

\begin{figure}[h]
\centering
\begin{tabular}{c c c}
  \begin{minipage}{4.8cm}
 \includegraphics[width=4.9cm]{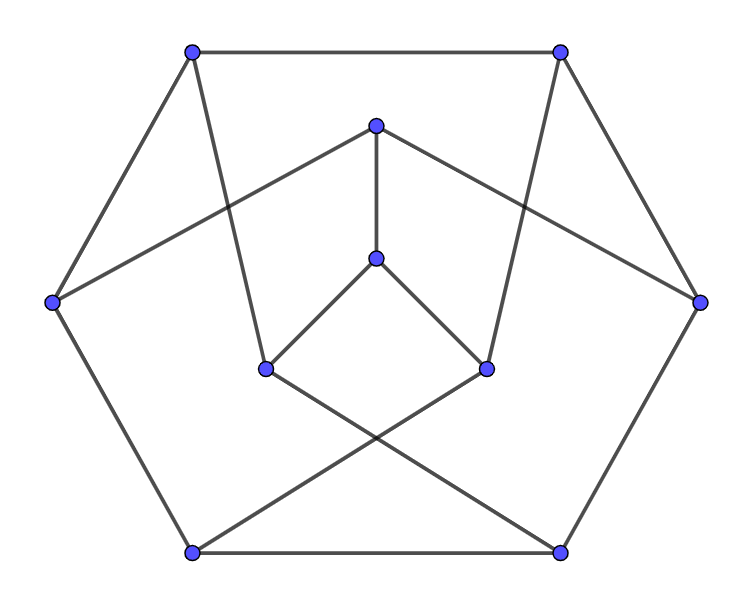}
  \end{minipage}
  &
  \begin{minipage}{4.8cm}
 \includegraphics[width=4.8cm]{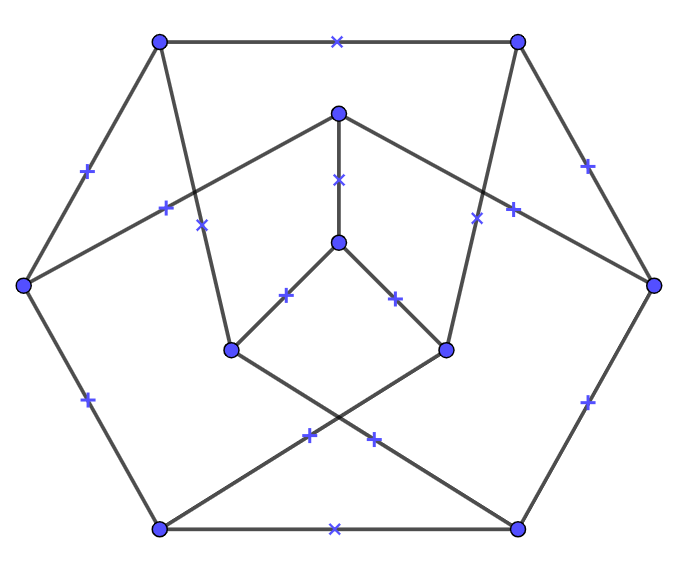}
  \end{minipage}
  &
  \begin{minipage}{4.8cm}
 \includegraphics[width=4.8cm]{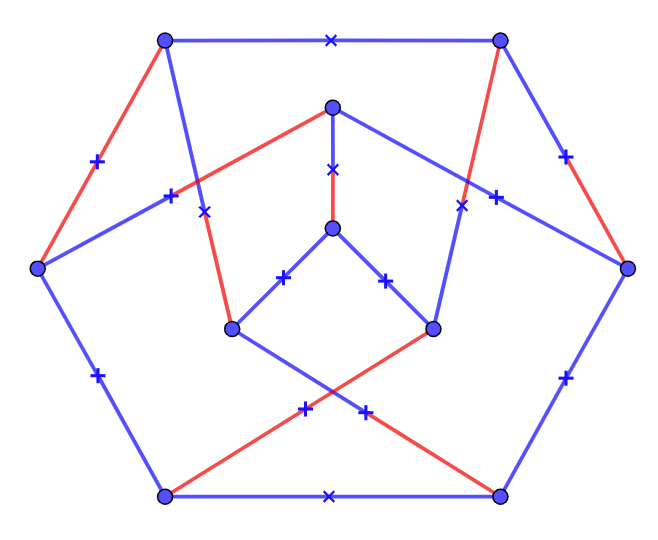}
  \end{minipage}
  \\
  \small (a) Petersen Graph
  &
  \small (b) Petersen Graph with
  &
  \small (c) A color configuration of\\
  & \small fictitious vertices
  & \small Petersen Graph
\end{tabular}
\caption{A complex coloring of Petersen Graph.}
\label{fig:01}
\end{figure}

\hspace{4mm}
The complex coloring is a variable elimination method. In a graph $G(V,E)$, suppose that a fictitious vertex is inserted in the middle of an edge $\langle {v_i,v_j} \rangle$ to divide the edge into two links. These two links connect the fictitious vertex to the two endpoints $v_i$ or $v_j$, respectively. As an example, the graph displayed in Figure~\ref{fig:01}(a) with inserted fictitious vertices is shown in Figure~\ref{fig:01}(b). Instead of coloring the edges, the complex coloring is assigning colors to links.

\begin{defn*}\emph{
Assigning the two colors $\{r:=red,~b:=blue\}$ to the two links of each edge, the coloring is \textbf{consistent} if one and only one of the links incident to each vertex is assigned red color $r$, and all other links are colored in blue color $b$. If the two links of an edge are colored with different colors $r$ and $b$, then the edge is called a $(r,b)$ \textbf{variable}, otherwise it is a \textbf{constant}, as Figure~\ref{fig:01} shows. The notation
$\langle {v_i,v_j} \rangle \rightarrow (color_1, color_2)$ is used to indicate the assigning of color pair $(color_1, color_2)$ to edge $\langle {v_i,v_j} \rangle $. The \textbf{color configuration} of a complex-colored graph $G$ is represented by the two-tuple $C(M)=\{M,\bar{\partial}(M)\}$, where $M$ is the set of edges that are fully colored in red color $r$, and $\bar{\partial}(M)$ is the set of vertices that are not covered by $M$ and they are incident to the red link of a $(r,b)$ variable. Vertices in $\bar{\partial}(M)$ are also called \textbf{$\boldsymbol{M}$-exposed}.
}
\end{defn*}

\hspace{4mm}
Since only one red link is incident to each vertex, there is a natural one-to-one correspondence between a matching $M$ and the color configuration $C(M)=\{M,\bar{\partial}(M)\}$, in which the set of red edges $M$ corresponds to a matching and $\bar{\partial}(M)$ is the set of free vertices relative to $M$. For example, the graph shown in Figure~\ref{fig:01}(c) is consistently colored by the set of colors $\{r,b\}$. The initial color assignment is random; the consistency requirement can be easily satisfied if we only assign the red color $r$ to one of the links incident to each vertex.

\begin{defn*}\emph{
The binary \textbf{color-exchange} operation $``\otimes"$ that operates on two adjacent colored edges $(color_1,color_2)$, $(color_3,color_4)$ is defined by,
\begin{align*}
    (color_1,color_2)\otimes(color_3,color_4):=
    (color_1,color_3)\odot(color_2,color_4)
\end{align*}
where $\odot$ indicates the adjacency of two-colored edges. A color exchange is \textbf{effective} if the operation does not increase the number of variables.
}
\end{defn*}

\begin{figure}[h]
\centering
\begin{tabular}{c c}
  \begin{minipage}{6.8cm}
 \includegraphics[width=6.0cm]{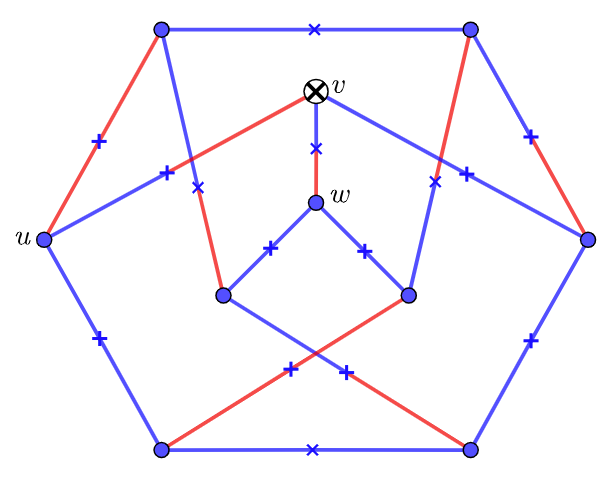}
  \end{minipage}
  &
  \begin{minipage}{6.8cm}
 \includegraphics[width=6.3cm]{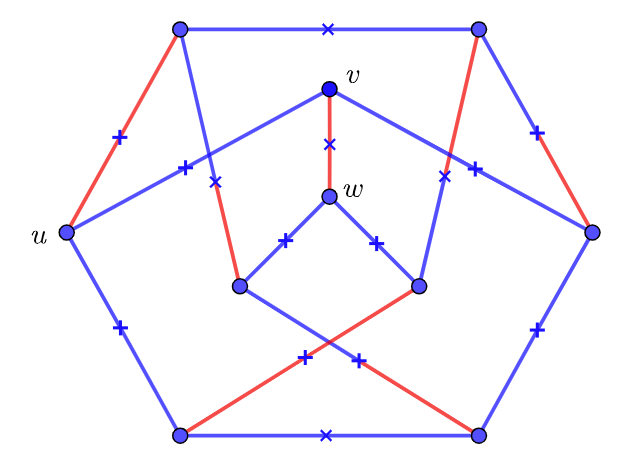}
  \end{minipage}
  \\
  \small (a) Before variable elimination
  &
  \small (b) After variable elimination
\end{tabular}
\caption{Variable elimination via color exchange operations.}
\label{fig:02}
\end{figure}

\hspace{4mm}
An example of the binary color-exchange operation $``\otimes"$ performed on two adjacent variables $\langle{u, v}\rangle\rightarrow (b,r)$ and $\langle{v, w}\rangle\rightarrow (b,r)$ is illustrated in Figure~\ref{fig:02}. The two variables were eliminated as the result of this color-exchange operation $(b,r)\otimes(b,r)=(b,b)\odot(r,r)$.

\hspace{4mm}
It is important to note that the above color-exchange operation preserves the consistency of a color configuration. Since we only allow effective color exchange operation $(color_1,color_2)\otimes(color_3,color_4)=
    (color_1,color_3)\odot(color_2,color_4)$
and the effectiveness is assured if either $(color_1,color_2)$ or $(color_3,color_4)$, or both are variable. Thus, the color exchange operation may either eliminate adjacent variables, or move a variable to an adjacent edge. Non-adjacent variables in a consistently colored graph must move next to each other before they can be eliminated.

\begin{figure}[h]
\centering
\begin{tabular}{c c}
  \begin{minipage}{6.8cm}
 \includegraphics[width=6.5cm]{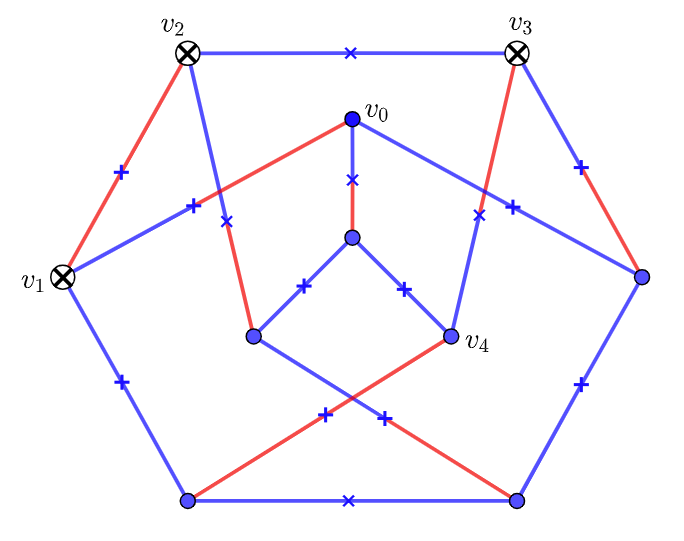}
  \end{minipage}
  &
  \begin{minipage}{6.8cm}
 \includegraphics[width=6.3cm]{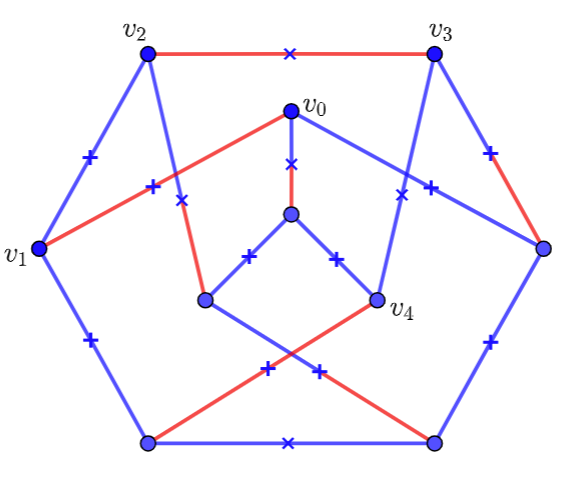}
  \end{minipage}
  \\
  \small (a) Before variable elimination
  &
  \small (b) After variable elimination
\end{tabular}
\caption{Variable elimination by walking on a two-colored augmenting path.}
\label{fig:03}
\end{figure}

\hspace{4mm}
Since a variable $(r,b)$ is always moving within an alternating path, the symmetric difference operation $M\oplus P$ defined in Lemma~\ref{lem:01} is equivalent to the elimination of two variables at the two ends of an augmenting path $P$.
As Figure~\ref{fig:03} shows, a $(r,b)$ variable $\langle{v_0,v_1}\rangle$ walks on a complex-colored augmenting path $\langle{v_0,v_1}\rangle, \langle{v_1,v_2}\rangle, \langle{v_2,v_3}\rangle$ by a sequence of color exchanges to cancel another $(r,b)$  variable $\langle{v_3,v_4}\rangle$. Note that the two end vertices $v_0$ and $v_3$ are both free vertices. Thus, according to Theorem~\ref{thm:01}, finding a maximum matching in a complex-colored graph $G$ is equivalent to repeatedly eliminating variables until remaining variables are all irreducible.

\section{Blossoms: Shrinking versus Deflection}\label{sec:03}

\begin{figure}[h]
\centering
\begin{tabular}{c}
  \begin{minipage}{12.8cm}
 \includegraphics[width=11.8cm]{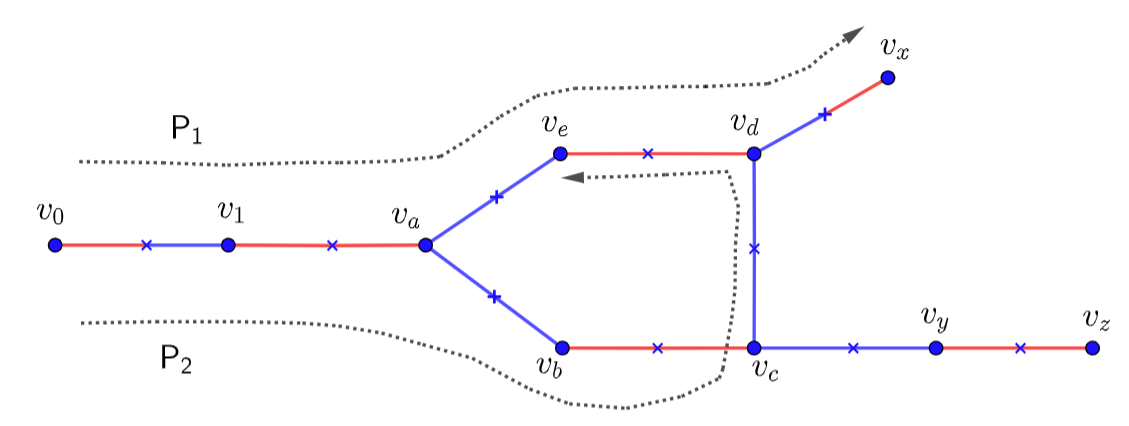}
  \end{minipage}
  \\
  \small (a) The original graph $G$\\
  \begin{minipage}{12.8cm}
 \includegraphics[width=11.8cm]{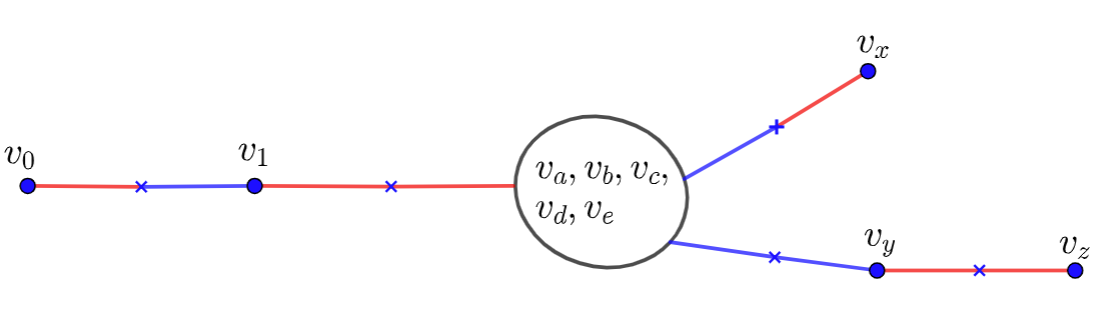}
  \end{minipage}
  \\
  \small (b) The contracted graph $G'$
\end{tabular}
\caption{Illustration of Blossom shrinking.}
\label{fig:04}
\end{figure}

\hspace{4mm}
The algorithm for maximum matching is a process of searching for successive $M$-augmenting paths starting from an initial matching $M$. In the exploration process of an $M$-augmenting path, the $M$-alternating path is a directional path, which starts from a free vertex and continuously grows in one direction.
In the matching $M$ depicted in Figure~\ref{fig:04}(a), there are two $M$-alternating paths starting from the free vertex $v_0$ that pass through $v_d$, namely
$$
P_1=v_0,v_1,v_a,v_e,v_d,v_x,
$$
and
$$
P_2=v_0,v_1,v_a,v_b,v_c,v_d,v_e.$$
If we take the $M$-alternating path $P_1$, then we can reach the other free vertex $v_x$ and obtain an $M$-augmenting path, in which the two end variables
$\langle{v_0,v_1}\rangle \rightarrow (r,b)$ and
$\langle{v_d,v_x}\rangle \rightarrow (b,r)$
can be eliminated by a sequence of color exchanges. However, if we take the $M$-alternating path $P_2$, then we miss this $M$-augmenting path. This divergent path problem arises when the vertex $v_d$ belongs to an odd cycle, called \textbf{blossom} by Edmonds.

\hspace{4mm}
Edmonds' algorithm solves this difficult problem by shrinking blossoms, or odd cycles, down to single super-vertices, thus to reveal augmenting paths, as Figure~\ref{fig:04}(b) shows. When we find an augmenting path from a free vertex $v_0$ to another free vertex $v_x$ in the contracted graph $G'$, then we immediately obtain an augment path in the original graph $G$ by unshrinking the super-vertices.

\begin{defn*}\emph{
Given a matching $M$ and an $M$-alternating path $P$ starting from a free vertex $v_0$, the \textbf{parity bit} of a vertex $v$ on $P$, denoted by $\pi(v)$, is determined by the distance (number of edges) between this vertex $v$ and the initial free vertex $v_0$ along $P$. If the distance is even then $\pi(v)=0$; otherwise, $\pi(v)=1$.
}
\end{defn*}

\hspace{4mm}
We adopt the convention that the link incident to the initial free vertex $v_0$ on an $M$-alternating path $P$ is always colored red, if not we can always change it to red by color-exchanging with its neighboring red link. With this convention, the parity bit of each intermediate vertex on the path $P$ can be defined by the following equivalent localized definition,
\begin{align*}
    \pi(v) =
    \left\{
    \begin{array}{ll}
    0,  &  \text{if input link to }v\text{ is red and output is blue (}v\text{ is in even state)}, \\
    1,     & \text{if input link to }v\text{ is blue and output is red  (}v\text{ is in odd state)}.
    \end{array}
    \right.
\end{align*}

\hspace{4mm}
Consider the two $M$-alternating paths,
\begin{align*}
    P_1 = v_0,v_1,v_a,v_e,v_d,v_c,v_b,
    \text{~~and~~}
    P_2 = v_0,v_1,v_a,v_b,v_c,v_d,v_e
\end{align*}
in the graph $G$ shown in Figure~\ref{fig:04}(a). We write the two paths $P_1$ and $P_2$ with their sequences of parities as follows:
\begin{align*}
&P_1:v_0 \pi(v_0 ) v_1 \pi(v_1 ) v_a \pi(v_a ) v_e \pi(v_e ) v_d \pi(v_d ) v_c \pi(v_c ) v_b \pi(v_b )=v_0 0v_1 1
{\color{magenta}
v_a 0}
{\color{blue}v_e 1v_d 0v_c 1v_b 0},\\
&P_2:v_0 \pi(v_0 ) v_1 \pi(v_1 ) v_a \pi(v_a ) v_b \pi(v_b ) v_c \pi(v_c ) v_d \pi(v_d ) v_e \pi(v_e )=v_0 0v_1 1
{\color{magenta}
v_a 0}
{\color{blue}v_b 1v_c 0v_d 1v_e 0}.
\end{align*}

Comparing the above two sequences, we can sum up the following properties of blossoms.

\begin{prop}\emph{
The $M$-alternating path $P$ always enters the blossom at a vertex $v_a$  with even parity bit $\pi(v_a )=0$, called \textbf{base}, because path divergence occurs only when the input link to $v_a$ is red and multiple outputs are blue. On the other hand, if the input link is blue, then there is only one red output, which is the case of entering the base of an even alternating cycle.
}
\end{prop}

\begin{prop}\emph{
The parity bit $\pi(v)$  of a vertex $v$ in the blossom, other than the base, can be either 0 (even) or 1 (odd), depending on the direction of the path $P$.
}
\end{prop}

\begin{prop}\emph{
If the $M$-alternating path $P$ return to the base $v_a$ and form a blossom, then the last vertex $v$ always possesses an even parity $\pi(v)=0$, which conflicts with the parity $\pi(v_a)=0$ of the base vertex $v_a$. For example, the last vertex $v_b$ in $P_1$, and $v_e$ in $P_2$.
}
\end{prop}

\hspace{4mm}
In the exploration of $M$-alternating paths, the difficulty arising from blossoms is mainly due the parity conflicts characterized in Property 2 and 3. The aim of shrinking the blossom to a single super-vertex is two-fold: eliminating the parity conflicts, and reserving all $M$-alternating paths passing through the blossom.

\hspace{4mm}
In contrast to shrinking, the algorithm proposed in this paper deflects the $M$-alternating path and makes a detour around blossoms. This dynamic exploration mechanism is a two tuple $T=\{P,S\}$, called \textbf{trunk}, which consists of an $M$-alternating path $P$ starting from a free vertex, and a stack of \textbf{sprout} $S$ that maintains all possible detours of $P$. The $M$-alternating path $P$ is a stack of ordered sequence of vertices, and the sprout $S$ is a stack of ordered sequence of edges, in which each edge is a sprout that represents the starting point of a reserved detour for the alternating path $P$.

\begin{defn*}\emph{
A vertex $v$ in the $M$-alternating path $P$ with \textbf{even parity} $\pi(v)=0$ is called a \textbf{sprout root} and is abbreviated as \textbf{$\boldsymbol{s}$-root}. The set of \textbf{free edges} incident with an $s$-root $v$ is defined as
$$
Sprout(v) = \left\{\langle v,u\rangle~|~v\in P,~\pi(v) = 0,~\langle{v},u\rangle\in\overline{M}\right\},
$$
and the set of vertices mated with an $s$-root $v$ by free edges is defined as
$$
Mate(v) = \left\{
u~|~v\in P,~\pi(v) = 0,~
\langle{v},u\rangle\in\overline{M}
\right\}.
$$
}
\end{defn*}

\hspace{4mm}
The $M$-alternating path $P$ is directional; at an odd parity matched vertex $v$ with $\pi(v)=1$, there is a unique path to continue $P$ from a blue input link to the only red output link. However, at an even matched parity vertex $v$ with $\pi(v)=0$, the vertex $v$ is an $s$-root, and the path $P$ can be continued from a red input link to any one of the multiple blue output links. In our DFS algorithm, the path $P$ will arbitrarily select one of the edges in $Sprout(v)$, and keep the others in reserve in the sprout stack $S$, in case that the path $P$ needs detours in the future.

\hspace{4mm}
The searching process of this dynamic trunk $T=\{P,S\}$ starts from an initial free vertex $v_0$ and one of its mate $u \in Mate(v_0 )$, meaning that initially we have $P=\{v_0,u\}$ with sprout set $S=Sprout(v_0 )\setminus\{\langle v_0,u\rangle\}$.
As the path $P$ extends, the process keeps adding pairs of vertices to the alternating path $P$, and appending sprouts to the stack $S$ along the extension of path $P$. If the path $P$ hits another free vertex, then an $M$-augmenting path is identified and the searching process stops. Otherwise, the exploration process will continue until the path hits a \textbf{dead end} or an \textbf{active vertex} in $P$. The latter case indicates that the path $P$ forms a cycle. In either case, the path $P$ will make a detour. The algorithm concedes defeat if the stack of sprout $S$ is empty, otherwise it will retrieve the last sprout in $S$, namely an edge $e=\langle v_s,v_t \rangle$, and replace the entire sub-path in $P$ starting from $v_s$ with the sequence $v_s,v_t$. The algorithm continues the searching process after making the detour. Table~\ref{table:01} lists each step of the searching process starting from the free vertex $v_0$ in the graph $G$ shown in Figure~\ref{fig:04}(a).

\begin{figure}[H]
\centering
\includegraphics[width=.6\linewidth]{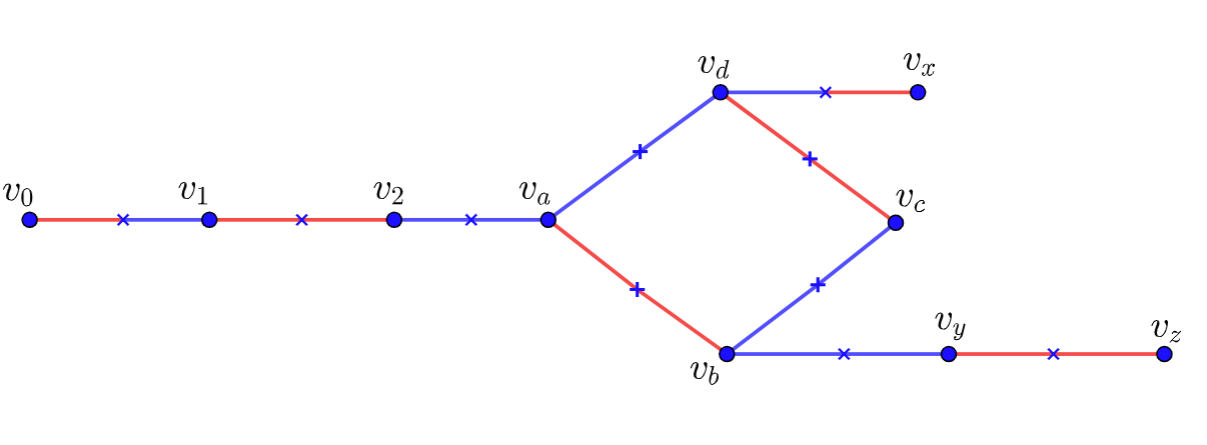}
\caption{The graph $G$ with an even cycle.}
\label{fig:05}
\end{figure}

\hspace{4mm}
The parity conflicts will not occur when the alternating path forms an even cycle. As Figure~\ref{fig:05} shows, there is only one $M$-alternating path transits the even cycle because the parity of the \textbf{base} vertex $v_a$ of the even cycle is odd with $\pi(v_a )=1$. Unlike odd cycles, an even cycle is a legitimate two-colored $M$-alternating cycle, which is naturally compatible with the $M$-alternating path $P$. The odd cycle and even cycle displayed in Figure~\ref{fig:04} and Figure~\ref{fig:05}, respectively, clearly demonstrate this key point. Table~\ref{table:02} provides the searching process starting from the free vertex $v_0$ in the graph shown in Figure~\ref{fig:05}.

\newpage
\newgeometry{left=5mm, right=5mm, bottom=26.5mm}

\begin{table}[H]
\centering
\caption{The process of searching for an augmenting path through an odd cycle.}
\label{table:01}
{\small
\begin{tabular}{|l|l|l|l|}\hline\hline
{\large Steps} &  {\large Alternating Path $P$} & {\large Sprout Stack $S$} & {\large Remarks}\\\hline\hline
\textbf{1} (Initialization) & ${v_0} 0{v_1} 1$ & $\emptyset$ & $v_0$ is the initial free vertex, add $v_0, v_1$ to $P$. \\\hline
\textbf{2} & ${v_0}0{v_1}1{v_a}0{v_b}1$ & $\langle v_a,v_e\rangle$ & Add $v_a, v_b$ to $P$ and $\langle v_a, v_e\rangle$ to $S$. \\\hline
\textbf{3} & ${v_0}0{v_1}1{v_a}0{v_b}1{v_c}0{v_d}1$ & $\langle v_a,v_e\rangle,\langle v_c,v_y\rangle$ & Add $v_c, v_d$ to $P$ and $\langle v_c,v_y\rangle$ to $S$. \\\hline
\textbf{4} & ${v_0}0{v_1}1{v_a}0{v_b}1{v_c}0{v_d}1{v_e}0{v_a}1$  &  $\langle v_a,v_e\rangle,\langle v_c,v_y\rangle$ &
\begin{tabular}{@{}l@{}}
Add $v_e, v_a$ to $P$, the vertex $v_a$ appeared\\
twice in $P$ with conflict parity, detect an\\
odd cycle.
\end{tabular}
\\\hline
\textbf{5} (Detour) & ${v_0}0{v_1}1{v_a}0{v_b}1{v_c}0{v_y}1$ & $\langle v_a,v_e\rangle$ &
\begin{tabular}{@{}l@{}}
Make a detour around cycle. \\
Retrieve sprout $\langle v_c,v_y\rangle$ from $S$, and replace\\
the sequence $v_c,v_d,v_e,v_a$ in $P$ with $v_c,v_y$.
\end{tabular}
\\\hline
\textbf{6} (Dead end) & ${v_0}0{v_1}1{v_a}0{v_b}1{v_c}0{v_y}1$ & $\langle v_a,v_e\rangle$  &
\begin{tabular}{@{}l@{}}
The path hits a dead end at $v_z$.
\end{tabular}
\\\hline
\textbf{7} (Detour) & ${v_0}0{v_1}1{v_a}0{v_e}1$ & $\emptyset$  &
\begin{tabular}{@{}l@{}}
Make a detour around the dead end $v_z$.\\
Retrieve sprout $\langle v_a,v_e\rangle$ from $S$, and replace\\
the sequence $v_a,v_b,v_c,v_y$ in $P$ with $v_a,v_e$.\\
Starting from here, the path is in the\\ clockwise
direction of the odd cycle\\ $v_a,v_b,v_c,v_d,v_e$.
\end{tabular}
\\\hline
\textbf{8} (Termination)& ${v_0}0{v_1}1{v_a}0{v_e}1{v_d}0{v_x}1$ &  $\langle v_d,v_c\rangle$ &
\begin{tabular}{@{}l@{}}
If $\langle v_d,v_x\rangle$ is selected, add $v_d,v_x$ to $P$, the\\ process
may move to $v_c$ or to $v_x$, in the \\latter case, the
augmenting path\\ $v_0,v_1,v_a,v_e,v_d,v_x$ is
identified and the \\process is stopped. \\
If $\langle v_d,v_c\rangle$ is selected then the next step is \textbf{8A}.
\end{tabular}
\\\hline
\textbf{8A} & ${v_0}0{v_1}1{v_a}0{v_e}1{v_d}0{v_c}1$ & $\langle v_d,v_x\rangle$  &
\begin{tabular}{@{}l@{}}
If the process selects $\langle v_d,v_c\rangle$ instead of\\
$\langle v_d,v_x\rangle$
in \textbf{step 8}, then add $v_d,v_c$ to $P$,\\
and $\langle v_d,v_x\rangle$ to $S$.
\end{tabular}
\\\hline
\textbf{9} & ${v_0}0{v_1}1{v_a}0{v_e}1{v_d}0{v_c}1{v_b}0{v_a}1$ & $\langle v_d,v_x\rangle$  &
\begin{tabular}{@{}l@{}}
Add $v_b,v_a$ to $P$, the vertex $v_a$ appeared twice\\
in $P$ with conflict parity, detect an odd cycle.
\end{tabular}
\\\hline
\begin{tabular}{@{}l@{}}
\textbf{10}
(Detour \\
and termination)
\end{tabular}
 & ${v_0}0{v_1}1{v_a}0{v_e}1{v_d}0{v_x}1$  & $\emptyset$ &
\begin{tabular}{@{}l@{}}
Make a detour around cycle. \\
Retrieve sprout $\langle v_d,v_x\rangle$ from $S$, and replace\\
the sequence $v_d,v_c,v_b,v_a$ in $P$ with $v_d,v_x$.\\
The augmenting path $v_0,v_1,v_a,v_e,v_d,v_x$ is\\
identified and the process is stopped.
\end{tabular}
\\\hline
\end{tabular}
}
\end{table}

\newpage

\begin{table}[H]
\centering
\caption{The process of searching for an augmenting path through an even cycle.}
\label{table:02}
{\small
\begin{tabular}{|l|l|l|l|}\hline\hline
  {\large Steps} &  {\large Alternating Path $P$} & {\large Sprout Stack $S$} & {\large Remarks}\\\hline\hline
   \textbf{1} (Initialization) & ${v_0} 0{v_1} 1$ & $\emptyset$ & $v_0$ is the initial free vertex, add $v_0, v_1$ to $P$. \\\hline
   \textbf{2}  & ${v_0} 0{v_1} 1{v_2} 0{v_a} 1$ & $\emptyset$ & Add $v_2, v_a$ to $P$. \\\hline
   \textbf{3}  & ${v_0} 0{v_1} 1{v_2} 0{v_a} 1{v_b} 0{v_c} 1$ & ${\langle v_b,v_y\rangle}$ &
   \begin{tabular}{@{}l@{}}
   If the edge ${\langle v_b,v_c \rangle}$ is selected, add $v_b, v_c$ to $P$ \\
   and ${\langle v_b,v_y \rangle}$  to $S$. \\
   Otherwise, the next step is \textbf{3A}.\end{tabular}
   \\\hline
   \textbf{4}  & ${v_0} 0{v_1} 1{v_2} 0{v_a} 1{v_b} 0{v_c} 1{v_d} 0{v_a} 1$ & ${\langle v_b,v_y\rangle}$, ${\langle v_d,v_x\rangle}$ &
   \begin{tabular}{@{}l@{}}
   Add $v_d, v_a$ to $P$, the vertex $v_a$ appeared twice \\
   in $P$ with same parity, detect an even cycle.
   \end{tabular}
   \\\hline
\begin{tabular}{@{}l@{}}
\textbf{5}
(Detour \\
and termination)
\end{tabular} & ${v_0} 0{v_1} 1{v_2} 0{v_a} 1{v_b} 0{v_c} 1{v_d} 0{v_x} 1$ & ${\langle v_b,v_y\rangle}$ &
   \begin{tabular}{@{}l@{}}
   Make a detour around the cycle. \\
   Retrieve sprout $\langle v_d,v_x \rangle$ from $S$, and replace \\
   the sequence $v_d, v_a$ in $P$ with $v_d, v_x$.\\
   The augmenting path $v_0,v_1,v_2,v_a,v_b,v_c,v_d,v_x$\\
   is identified and the process is stopped.
   \end{tabular}
   \\\hline
   \textbf{3A} (Dead end) & ${v_0} 0{v_1} 1{v_2} 0{v_a} 1{v_b} 0{v_y} 1$ & ${\langle v_b,v_c\rangle}$ &
   \begin{tabular}{@{}l@{}}
   If the edge $\langle v_b,v_y \rangle$ is selected, add $v_b, v_y$ to $P$ \\
   and $\langle v_b,v_c \rangle$  to $S$. \\
   Then the path $P$ hits a dead end at $v_z$.
   \end{tabular}
   \\\hline
   \textbf{4A} (Detour) & ${v_0} 0{v_1} 1{v_2} 0{v_a} 1{v_b} 0{v_c} 1$ & $\emptyset$ &
   \begin{tabular}{@{}l@{}}
   Make a detour around the dead end $v_z$. \\
   Retrieve sprout $\langle v_b,v_c \rangle$ from $S$, and replace the\\ sequence $v_b, v_y$ in $P$ with $v_b, v_c$.

   \end{tabular}
   \\\hline
   \textbf{5A} (Termination) & ${v_0} 0{v_1} 1{v_2} 0{v_a} 1{v_b} 0{v_c} 1{v_d} 0{v_x} 1$ & ${\langle v_d,v_a\rangle}$ &
   \begin{tabular}{@{}l@{}}
   Add $v_d, v_x$ to $P$ and $\langle v_d,v_a \rangle$ to $S$.\\
   The augmenting path $v_0,v_1,v_2,v_a,v_b,v_c,v_d,v_x$ \\
   is identified and the process is stopped.
   \end{tabular}
   \\\hline
\end{tabular}
}
\end{table}

\restoregeometry


\begin{figure}[ht]
\centering
\includegraphics[width=.55\linewidth]{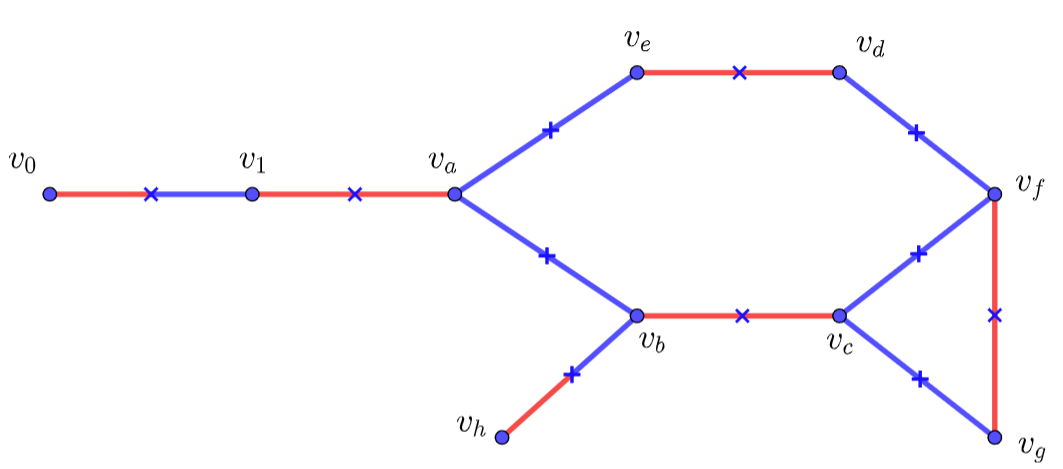}
\caption{The graph $G$ with two nested odd-cycles.}
\label{fig:08}
\end{figure}

\hspace{4mm}
The searching process adaptively changes the directional alternating path according to the topology of the graph. The graph $G$ displayed in Figure~\ref{fig:08} has two nested odd cycles. Starting at free vertex $v_0$, the alternating path $P$ encountered odd cycles four times before it finds another free vertex $v_h$. The following sequence of the searching process reveals the resilience of the dynamic trunk $T=\left\{P,S\right\}$.
\begin{enumerate}
\item [1).] 	$P=v_0{0}v_1{1}v_a{0}v_b{1}v_c{0}v_f{1}
v_g{0}v_c{1}$\\
$S=\left\{\langle{v_a,v_e}\rangle,\langle{v_c,v_g}\rangle\right\}$

\item [2).]
$P=v_0{0}v_1{1}v_a{0}v_b{1}{v_c}0{v_g}1$\\
$S=\left\{\langle{v_a,v_e}\rangle\right\}$

\item [3).]
$P=v_0{0}v_1{1}v_a{0}v_b{1}{v_c}0{v_g}1{v_f}0{v_d}1{v_e}0{v_a}1$\\
$S=\left\{
\langle{v_a,v_e}\rangle,
\langle{v_f,v_c}\rangle
\right\}$

\item [4).]
$P=v_0{0}v_1{1}v_a{0}v_b{1}{v_c}0{v_g}1{v_f}0{v_c}1$\\
$S=\left\{
\langle{v_a,v_e}\rangle
\right\}$

\item [5).]
$P=v_0{0}v_1{1}v_a{0}v_e{1}$\\
$S=\emptyset$

\item [6).]
$P=v_0{0}v_1{1}v_a{0}v_e{1}v_d{0}v_f{1}
v_g{0}v_c{1}
v_b{0}v_a{1}$\\
$S=\left\{
\langle{v_b,v_h}\rangle
\right\}$

\item [7).]
$P=v_0{0}v_1{1}v_a{0}v_e{1}v_d{0}v_f{1}v_g{0}v_c{1}v_b{0}v_h{1}$\\
$S=\emptyset$\\
$/{^{*}}$ augmenting path identified successfully.$^{*}\hspace{-1mm}/$
\end{enumerate}

\section{The DFS Algorithm}\label{sec:04}
\hspace{4mm}
In this Section, we summarize the DFS algorithm and describe the details of the searching process. The algorithm consists of two phases: a \emph{growing phase} and a \emph{pruning phase}. The alternating path $P$ and the sprout stack $S$ will be updated in both phases.

\textbf{Input}: A general graph $G(V,E)$, a color configuration $C(M)=\{M,\bar{\partial}(M)\}$ of graph $G(V,E)$ with a current matching $M$, and a free vertex $v_0\in \bar{\partial}(M)$.

\textbf{Idea}: Explore a trunk $T=\{P,S\}$ from the $M$-exposed vertex $v_0$, stretching the alternating path $P$ and the sprout stack $S$ as far as possible. In each step, an ordered pair of vertices $(v_a,v_b)$ will be added to the alternating path $P$; the first vertex $v_a$ is an $s$-root with even parity in $P$. The edge $e=\langle v_a, v_b \rangle$ is a sprout selected from the set $Sprout(v_a)$; the rest edges in $Sprout(v_a )\setminus \{\langle v_a, v_b\rangle \}$ will then be added to $S$.
If a dead end was detected or a cycle was formed,
the trunk $T$ will make a detour according to the last sprout in stack $S$.
Declare a failure if $S$ is empty, otherwise continue the process to reach another free vertex that yields an augmentation.

\textbf{Initialization}: $P=\{v_0,v_1 \}$,
$S=Sprout(v_0 )\setminus\{\langle v_0, v_1 \rangle\}$.

\textbf{Iteration}: (\textbf{Growing Phase}) If the next pair of vertices $\langle v_a, v_b\rangle$  extended from the current alternating path $P$ are not in $P$, then perform the following updating operation:
$$
P=P\cup \{v_a, v_b \},
$$
$$
S=S\cup Sprout(v_a )\setminus\{\langle v_a,v_b\rangle \}.
$$
(\textbf{Pruning Phase})
When a dead end or a cycle was found, stop if $S=\emptyset$ and there is no $M$-augmenting path starting from $v_0$, otherwise,
retrieve the last sprout $\langle v_a,v_b \rangle$ from $S$, eliminate all vertices after $v_a$ in the path $P=v_0,\ldots,v_a,\ldots,v_x $
and replace them with $v_a,v_b$.
Update trunk $T$ as follows and continue the searching process:
$$
P=v_0,\ldots,v_a,v_b,
$$
$$
S=S\setminus \{\langle v_a,v_b \rangle\}.
$$

\hspace{4mm}
In the pruning phase, we implicitly claim that if the edge $\langle v_a,v_b\rangle \in S$
then the vertex $v_a\in P$. This claim is always valid because the vertex $v_a$ and the edge $\langle v_a,v_b\rangle \in Sprout(v_a)$ were added to $P$ and $S$, respectively and simultaneously, in the growing phase.

\hspace{4mm}
The algorithm terminates when no augmenting paths exist. The searching process halts when either an augmenting path was identified, or every alternating path starting from a free vertex was inspected and returned with an empty sprout stack. We show in the following lemma that all possible alternating paths starting from a free vertex $v_0$ will be visited if the exploration process
ends with an empty sprout stack.

\begin{lem}\label{lem:02}
{\emph{
If an alternating path $P$ starting from a free vertex $v_0$ ends the searching process with an empty sprout stack $S=\emptyset$, then $P$ has visited every alternating path starting from $v_0$.}
}
\end{lem}
\vspace{-3mm}
\proof{}
Suppose $Q=v_0,\cdots,v_x,v_a,v_b$  is the shortest alternating path that $P$ has never visited, where $\pi\left(v_a\right)=0$ and $\pi\left(v_b\right)=1$.
Then neither $P$ has visited the alternating path $Q^{\prime}=v_0,\cdots,v_x$, because $\langle{v_x,v_a}\rangle \in Sprout(v_x)$ but it was eventually disappeared in the final sprout stack $S=\emptyset$. That is, if $P$ has visited $Q^{\prime}=v_0,\cdots,v_x$ then it certainly has visited $Q=v_0,\cdots,v_x,v_a,v_b$ through the sprout $\langle{v_x,v_a}\rangle$, which is impossible according to our assumption.
On the other hand, if $P$ has never visited $Q^{\prime}=v_0,\cdots,v_x$, then this contradicts our assumption that $Q=v_0,\cdots,v_x,v_a,v_b$ is the shortest alternating path that $P$ has never visited.
\endproof{}

\hspace{4mm}
In the DFS algorithm, we assume that if a free vertex $v_0$ failed to find another free vertex through an alternating path, then this $v_0$ will never access any other free vertices, even if other augmenting paths modified the graph configuration. By definition, any alternating path in a maximum matching $M$ should contain at most one $M$-exposed vertex. This point can be further elaborated by the Gallai-Edmonds decomposition of a graph $G$, in which every $M$-exposed vertex $v_0$ of a maximum matching $M$ is locked up in an odd component of $G$. This isolation property ensures that repeating an exploring process starting from the same free vertex $v_0$ is not necessary.

\begin{figure}[ht]
\centering
\includegraphics[width=.40\linewidth]{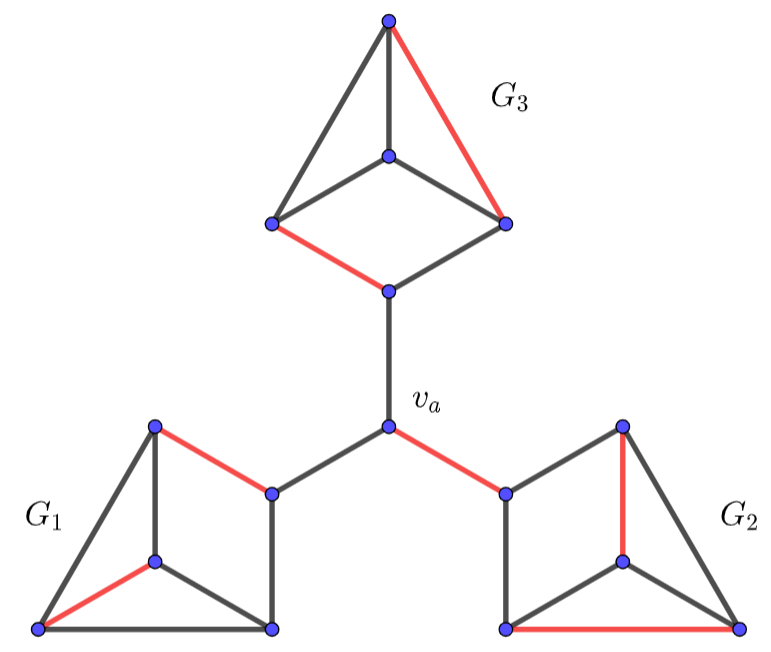}
\caption{A maximum matching of the Sylvester's graph.}
\label{fig:09}
\end{figure}

\hspace{4mm}
The Sylvester's graph is a good example to illustrate the isolation property of free vertices in a maximum matching $M$. As Figure~\ref{fig:09} shows, the three odd components $G_1$, $G_2$, $G_3$ of graph $G$ are connected by a vertex $v_a$, we observe the following properties:
\vspace{-3mm}
\begin{enumerate}
\item [1).]	
Deleting $v_a$, $M$ covers all but one vertex of each odd component $G_i$, $i=1,2,3$.

\item [2).]
$M$ covers the vertex $v_a$.

\item [3).] 	
If $M$ matches one of the free vertices in $G_i$, $i=1,2,3$, with $v_a$, then the other two free vertices in $G_j$, $j\neq i$, will be isolated, and they cannot be connected by an alternating path.
\end{enumerate}

\vspace{-3mm}

\begin{defn*}
{\emph{
In a graph $G\left(V,E\right)$, for $S\subseteq V\left(G\right)$, let $N_G\left(S\right)$ denote the set of vertices in $G-S$ which have at least one neighbor in $S$, and let $G\left[S\right]$ denote the subgraph of $G$ induced by $S$. The graph $G$ is \textbf{factor-critical} if $G-v$ has a perfect matching for every vertex $v\in V\left(G\right)$. A matching in $G$ is \textbf{near-perfect} if it matches all but one vertex of $G$.
}}
\end{defn*}

\hspace{4mm}
A factor-critical graph is connected, and has an odd number of vertices. Simple examples include odd-length cycle $C_n$ and the complete graph $K_n$ of odd order $n$.

\begin{defn*}
{\emph{
In a graph $G\left(V,E\right)$, let $B$ be the set of vertices covered by every maximum matching in $G$, and let $D=V\left(G\right)-B$. The set B is further partitioned into $B=A\cup C$, where $A$ is the set of vertices that are adjacent to at least one vertex in $D$, and $C=B-A$. The Gallai-Edmonds decomposition of $G$ is the partition of $V\left(G\right)$ into three
mutually disjoint
subsets $V=A\cup C\cup D$.
}}
\end{defn*}

\begin{thm}[Gallai-Edmonds Structure Theorem \cite{gallai1963kritische}]\label{thm:03}
{\emph{
Let $A$, $C$, $D$ be the sets in the Gallai-Edmonds Decomposition of a graph $G\left(V,E\right)$. Let $T_1, \cdots, T_l$ be the components of $G\left[C\right]$, and $G_1, \cdots, G_k$ be the components of $G\left[D\right]$. If $M$ is a maximum matching in $G$, then the following properties hold:
\vspace{-3mm}
\begin{enumerate}
\item [1).]  Each $T_i$, $i=1,2,\ldots, l$ is an even component, and $M$ restricts to a perfect matching on $T_i$.
\item [2).] Each $G_i$, $i=1,2,\ldots, k$ is an odd component, which is factor-critical, and $M$ restricts to a near-perfect matching on $G_i$.
\item [3).] $M$ completely matches $A$ into distinct components $G_1, \cdots, G_k$ of $G\left[D\right]$.
\end{enumerate}
}}
\end{thm}

\vspace{-2mm}
\hspace{4mm}
A detailed proof of this theorem is given in \cite{lovasz1986matching}, and a short proof is provided in \cite{west2011short}.
The property 3 in Theorem~\ref{thm:03} can be explained by Hall's Theorem.
Contracting each component $G_i$ of $G\left[D\right]$ to a single vertex $v_{g,i}$, we define an auxiliary bipartite graph $H\left(A\cup Y,E_H\right)$ as follows:
\begin{align*}
\left\{
\begin{array}{l}
  Y=\left\{v_{g,1}, \cdots, v_{g,k}\right\},~\text{and}~A=\left\{v_{a,1}, \cdots, v_{a,h}\right\},\\
  E_H=\left\{
\left(v_{a,j},v_{g,i}\right)~|~  v_{a,j}\in A\text{  having a neighbor in }G_i
\right\}.
\end{array}
\right.
\end{align*}

\vspace{-5mm}

\begin{figure}[h]
\centering
\begin{tabular}{c}
  \begin{minipage}{9.8cm}
 \includegraphics[width=9.5cm]{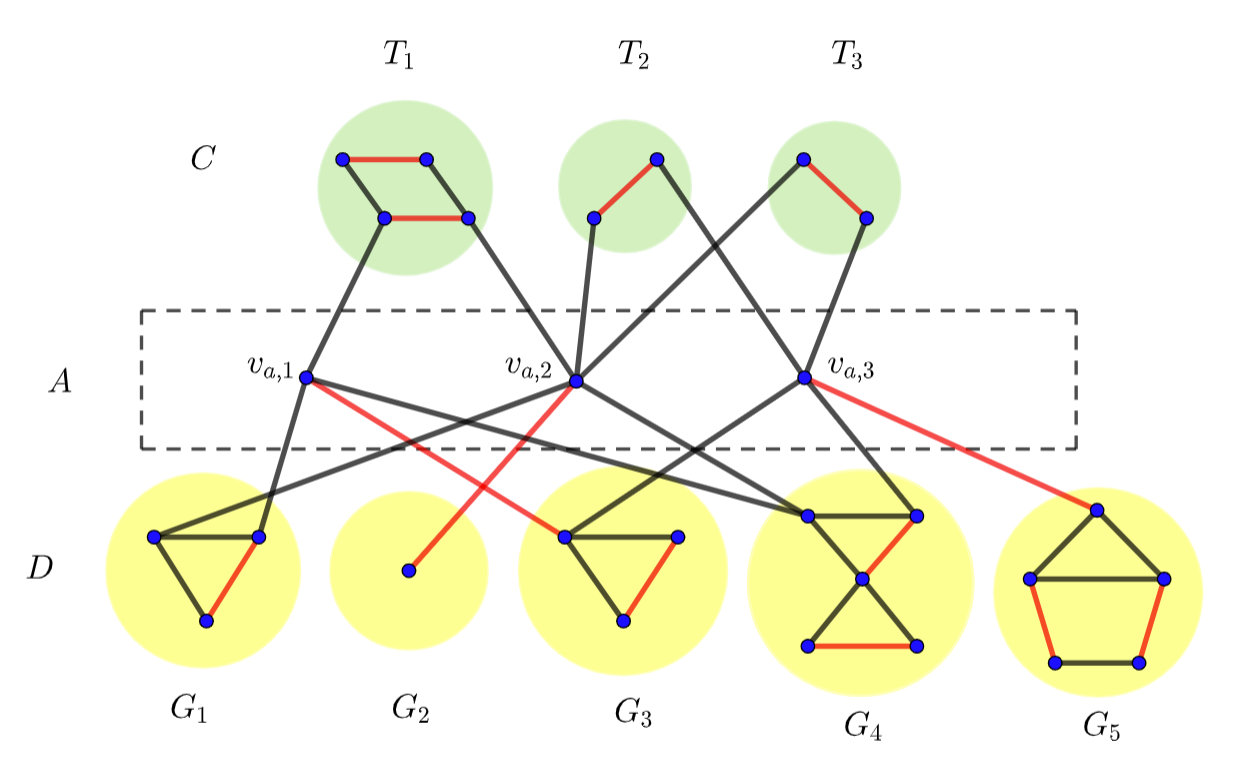}
  \end{minipage}
  \\
  \small (a) The decomposition $V=A\cup C\cup D$\\\vspace{5mm}
  \begin{minipage}{9.8cm}
 \includegraphics[width=9.5cm]{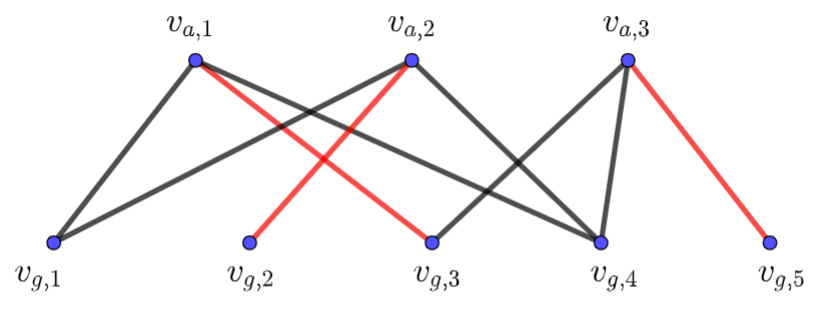}
  \end{minipage}
  \\
  \small (b) Auxiliary bipartite graph $H$
\end{tabular}
\caption{The Gallai-Edmonds decomposition of a general graph $G$.}
\label{fig:12}
\end{figure}

In the Gallai-Edmonds decomposition and a maximum matching $M$ of a general graph $G$, as Figure~\ref{fig:12} illustrates, the isolated $M$-exposed vertex in $G_1$ cannot access to that in $G_4$ by any alternating paths. It can be shown that
Hall's condition $|S|\leq |N_H(S)|$ holds for any
$S\subseteq A$ \cite{west2011short, west2001introduction},
thus the $M$ restricts to a matching on bipartite graph $H$ that covers $A$.

\hspace{4mm}
Since each odd component $G_i$ is factor-critical, and any vertex of $G_i$ can be the one unmatched by a maximum matching $M$. Thus, the only unmatched vertex in each odd component $G_i$ can either be matched with a vertex $v_a\in A$, or be isolated in the odd component $G_i$. Therefore, any free vertex can only be the source of an augmenting path at most once in the DFS algorithm. Since the initial number of free vertices is upper bounded by the order of $O(n)$, and the length of each alternating path $P$ is proportional to the number of edges $m=|E|$, the complexity of the DFS algorithm is given in Theorem~\ref{thm:04}.

\begin{thm}\label{thm:04}\emph{
The DFS algorithm can determine a maximum matching of a general graph in $O(mn)$
time with space complexity $O(n)$.
}
\end{thm}

\begin{figure}[ht]
\centering
\includegraphics[width=0.65\linewidth]{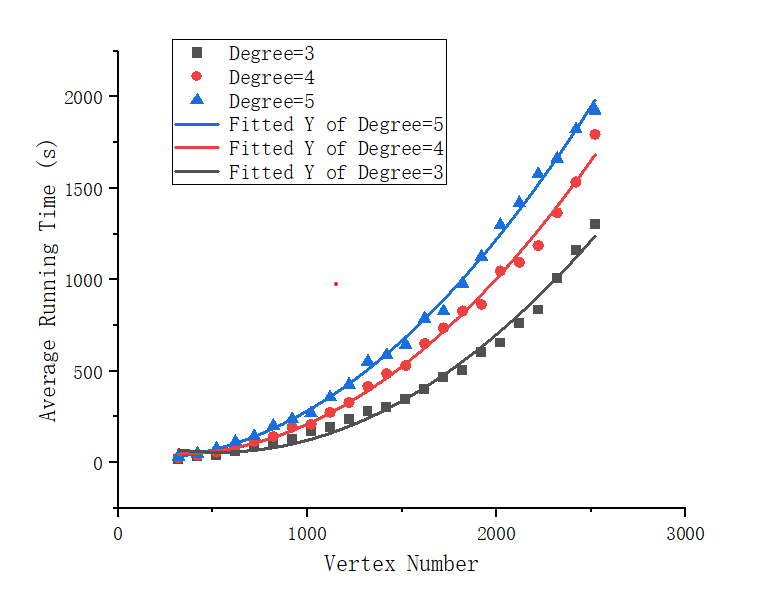}
\caption{The running time of the maximum matching algorithm.}
\label{fig:11}
\end{figure}

\hspace{4mm}
Experiments were conducted to verify the performance of our maximum matching algorithm. A set of $\Delta$-regular graphs with $n$ vertices and $m = \dfrac{\Delta{n}}{2}$ edges was randomly generated. Figure~\ref{fig:11} shows the experimental results of average running time, in which 25 graphs were randomly generated for every pair of $(\Delta,n)$, $n=100,200,\ldots,2500$ and $\Delta=3,4,5$.
These experimental results confirm the performance of our maximum matching algorithm given in Theorem~\ref{thm:04}. As shown in Figure~\ref{fig:11}, for each set of graphs under consideration, the running time of our algorithm is on the order of $O(n^2)$ for a given degree $\Delta$.

\section{Conclusion}\label{sec:conclusion}
\hspace{4mm}
The fundamental problem of finding maximum matchings in general graphs is the existence of odd cycles, or blossoms. Instead of shrinking blossoms, this paper proposed a deflection algorithm to cope with the parity conflicts caused by odd cycles. This new algorithm achieves $O(mn)$ time complexity with $O(n)$ space complexity.
This newly proposed algorithm is  complementary to Edmonds' blossom algorithm in two important aspects: depth-first search (DFS) versus breadth-first search (BFS), and deflection from blossoms versus shrinking of blossoms. In the future, we will explore the application of this method to maximum matching of weighted graphs.

\vspace{5mm}
\begin{center}
    {\large{
    ACKNOWLEDGMENTS
    }}
\end{center}

The authors would like to thank Professor Shahbaz Khan of Department of Computer Science and Engineering, Indian Institute of Technology, Roorkee, India, for many useful criticism and suggestions.

\bibliographystyle{ieeetr}

\bibliography{main}

\end{document}